\documentclass[%
 reprint,
superscriptaddress,
amsmath,
amssymb,
aps,
prl,
longbibliography
]{revtex4-2}

\usepackage{placeins}
\usepackage{xspace}
\usepackage{appendix}
\usepackage{graphicx}% Include figure files
\usepackage{dcolumn}% Align table columns on decimal point
\usepackage{bm}% bold math
\usepackage{booktabs}
\usepackage{hyperref}% add hypertext capabilities
\usepackage{xcolor}
\hypersetup{
    colorlinks,
    linkcolor={blue!70!black},
    citecolor={blue!70!black},
    urlcolor={blue!70!black}
}

\newcommand{\red}{\textcolor{black} }

\newcommand*{\DEstage}{\ensuremath{\Delta E_\text{stage}}\xspace}
\newcommand*{\Bmax}{\ensuremath{B_\text{max}}\xspace}
\newcommand*{\myB}{\ensuremath{B}}
\newcommand*{\muB}{\ensuremath{\mu_B}\xspace}
\newcommand*{\mysim}{\ensuremath{\sim \!}}
\newcommand*{\mygtr}{\ensuremath{> \hspace{-0.2em}}}

\newcommand*{\XSigma}{\ensuremath{\tilde{X}\,^2\Sigma^+}}
\newcommand*{\APiHalf}{\ensuremath{\tilde{A}\,^2\Pi_{1/2}}}
\newcommand*{\BSigma}{\ensuremath{\tilde{B}\,^2\Sigma^+}}

\begin{document}

\title{Zeeman-Sisyphus Deceleration of Molecular Beams}% Force line breaks with \\
\author{Benjamin L. Augenbraun}
\email{augenbraun@g.harvard.edu}
\author{Alexander Frenett}
\author{Hiromitsu Sawaoka}
\author{Christian Hallas}
\author{Nathaniel B. Vilas}
\author{Abdullah Nasir}
\author{Zack D. Lasner}
\author{John M. Doyle}
\affiliation{Department of Physics, Harvard University, Cambridge, MA 02138, USA}
\affiliation{Harvard-MIT Center for Ultracold Atoms, Cambridge, MA 02138, USA}

\date{September 7, 2021}% It is always \today, today,

\begin{abstract}
\noindent We present a robust, continuous molecular decelerator that employs high magnetic fields and few optical pumping steps. CaOH molecules are slowed, accumulating at low velocities in a range sufficient for loading both magnetic and magneto-optical traps. During the slowing, the molecules scatter only 7 photons, removing around 8~K of energy. Because large energies can be removed with only a few spontaneous radiative decays, this method can \red{in principle} be applied to nearly any paramagnetic atomic or molecular species, opening a general path to trapping of complex molecules.
\end{abstract}

\maketitle

Ultracold molecules are a very promising platform for applications spanning quantum simulation and quantum computation~\cite{Carr2009, demille2002quantum}, ultracold chemistry~\cite{krems2010viewpoint,balakrishnan2016perspective}, and precision searches for new particles~\cite{demille2015diatomic, Cairncross2019}. Molecules are also of interest  because their internal structures offer new handles for quantum control and sensing~\cite{bohn2017cold, demille2017probing}. For example, the rotation-vibration modes of small polyatomic molecules have been identified as offering orders-of-magnitude enhancement in sensitivity to the electric dipole moment of the electron~\cite{kozyryev2017PolyEDM, Augenbraun2021Observation} and potentially strong coupling to ultralight bosonic dark matter~\cite{kozyryev2021Enhanced}. Large polyatomic molecules open up additional opportunities in physics and chemistry~\cite{Wall2013,Yu2019, Klos2020, Augenbraun2020ATM, Yu2021Probing}. For instance, asymmetric top molecules provide new tools for quantum computation and simulation~\cite{albert2019robust, Dickerson2021} and controlling such species will enable precise studies of chiral molecules~\cite{quack1989structure, Quack2002}. These diverse applications share a dependence on the long coherence times and the exquisite quantum control achievable at ultralow temperature, a regime that is accessible by sub-Doppler laser cooling.

The first molecular laser cooling was demonstrated with diatomic molecules~\cite{Barry2014, truppe2017CaF, Collopy2018}. Key advances that led to this relied on rapid photon cycling, a process in which molecules repeatedly scatter many thousands of optical photons while avoiding loss to ``dark" rotational or vibrational sublevels. Laser cooling of polyatomic molecules, including linear triatomic molecules~\cite{kozyryev2016Sisyphus, AugenbraunYbOHSisyphus, Baum2020} and nonlinear symmetric top molecules~\cite{mitra2020direct}, has been achieved in one dimension, similar to early seminal work with diatomic molecules~\cite{Shuman2010}. Extension to three-dimensional cooling and trapping is challenging because the number of potential dark states increases in proportion to the size of a polyatomic molecule. A central difficulty arises in the case of slowing polyatomic molecules from the velocities at which they are produced (typically $\gtrsim80$~m/s) to those at which they may be captured into a trap, for example a magneto-optical trap (MOT; $v_\text{trap}\mysim10$~m/s). 

All molecular MOTs to date have been loaded by decelerating diatomic molecular beams to trappable velocities using the radiation pressure force exerted by high-power laser beams~\cite{Hemmerling2016, Truppe2017b, Yeo2015}, either white-light or chirped slowing. Because of the small momentum change caused by each photon absorption-emission cycle, such radiative slowing requires scattering $n_\gamma \sim 10^4$ photons to reach $v_\text{trap}$. This $n_\gamma$ is at least an order of magnitude larger than the number of scattered photons that has so far been demonstrated for any polyatomic molecule~\cite{Baum2021}. Radiative forces are especially inefficient for decelerating significant numbers of large polyatomic species that are both massive and limited (either technically or intrinsically) to optically cycle only $\mysim100$s of photons~\cite{Dickerson2021, Dickerson2021Optical}. To circumvent this, a number of non-radiative slowing methods have been developed~\cite{Hudson2004, Fulton2004, Wall2011, LavertOfir2011, Akerman2017, liu2017magnetic}, but these are inherently pulsed, slowing ``packets" of molecules $\lesssim \! 100$~$\mu$s in duration.  Slowing of continuous molecular beams was achieved with a ``cryofuge" decelerator in work complementary to that presented here~\cite{Chervenkov2014continuous, Wu2017cryofuge}.

\begin{figure*}[tb]
    \centering 
    \includegraphics[width=1\textwidth]{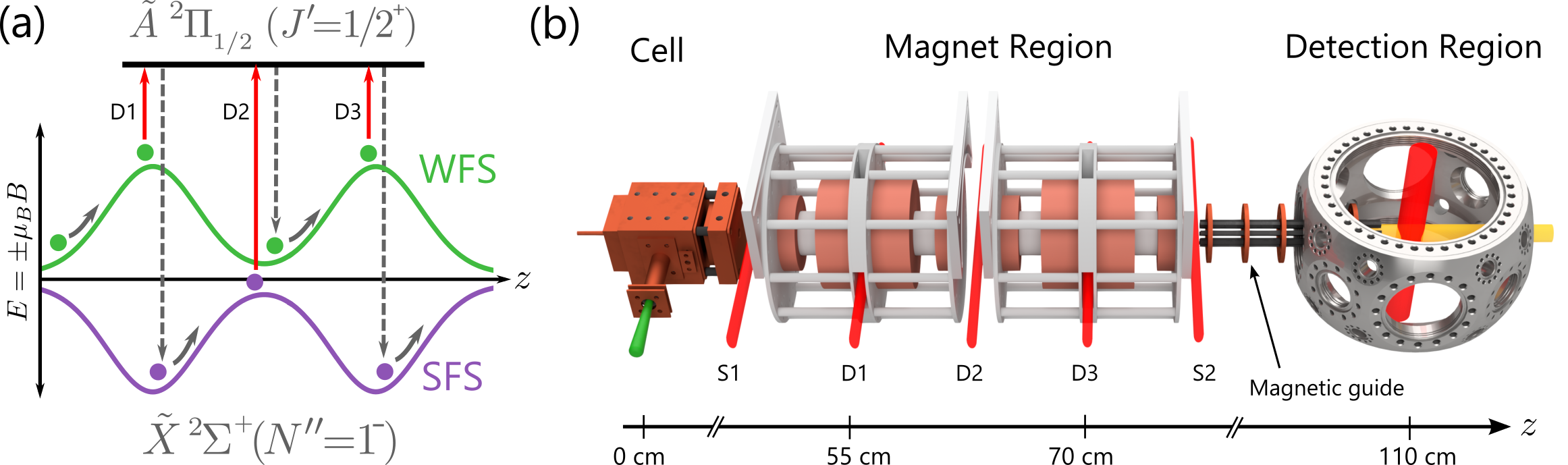}
    \caption{(a) Overview of the deceleration scheme. Molecules enter the magnetic field region in a weak-field-seeking state and decelerate as they travel toward the peak magnetic field. At the peak magnetic field, molecules are optically pumped to a strong-field-seeking (SFS) state and continue to decelerate. Near the field minimum, molecules are pumped back to the weak-field-seeking (WFS) state and the process can be repeated for additional deceleration. (b) Schematic of the experimental setup (not to scale). Molecules are produced in a two-stage cryogenic buffer-gas beam. They travel through two superconducting magnets and are optically pumped in three deceleration pumping regions (D1, D2, D3) by transverse laser beams at 626~nm. State-preparation regions (S1 and S2) pump molecules into WFS states in order to enhance the slowable population and populate magnetically guided states. Molecules are detected via laser-induced fluorescence by simultaneous excitation at 626~nm and 574~nm in a Doppler-sensitive configuration.}
    \label{fig:Setup}
\end{figure*}

In this Letter, we present a superconducting magnetic/optical decelerator that slows molecular beams to trappable velocities while using fewer than 10 spontaneous photon scatters. Deceleration is achieved by a combination of two large, static magnetic field regions and three electron-spin-flip optical transitions. We demonstrate this ``Zeeman-Sisyphus" (Z-S) decelerator~\cite{Fitch2016, Comparat2014Molecular} using a beam of the polyatomic molecule calcium monohydroxide (CaOH), an archetypal example of a broad class of laser-coolable polyatomic (and diatomic) molecules. We observe large accumulation of CaOH molecules at velocities below 15~m/s, with the average slowed molecule having scattered a low number of photons, $n_\gamma \sim 7$, to remove around 8~K of kinetic energy.  These results are in excellent agreement with numerical simulations, indicating that the deceleration process is very well understood.  The energy removal capability of this decelerator depends only on ground-state Zeeman shifts and is independent of the target species' mass. This establishes a general method to produce trappable samples of an entire class of molecules, including complex polyatomic molecules~\cite{Dickerson2021, kozyryev2016MOR, Augenbraun2020ATM}.

To produce this decelerator, we leverage the large energy shifts induced by Tesla-scale magnetic fields. The principle of our decelerator is depicted in Fig.~\ref{fig:Setup}(a). This scheme is motivated by the previous loading of a deep superconducting trap with CaF, as well as a proposed decelerator comprised of many stages of permanent magnets~\cite{Comparat2014Molecular, Fitch2016, lu2014magnetic}. The slowing process starts as molecules in a cryogenic buffer-gas beam (CBGB)~\cite{hutzler2012buffer} in a weak-field-seeking (WFS) state are incident on the fringing magnetic field of a compensated solenoid. As the molecules traverse the region of increasing magnetic field magnitude, they decelerate. (The magnetic tuning of CaOH energy levels relevant to this work is discussed in detail in \footnote{\label{note1}See Supplemental Material for additional details.}.) Near the magnetic field maximum [at location D1 in Fig.~\ref{fig:Setup}(b)], the molecules are optically pumped through an electronically excited state to a strong-field-seeking (SFS) state and continue to decelerate as they exit the high-field region. In this way, an energy $\DEstage \approx 2 \muB \Bmax$ can be removed from molecules passing through each deceleration stage, where \muB is the Bohr magneton and $\Bmax$ is the maximum magnetic field in the high-field region. This process can be repeated to remove additional energy. Furthermore, the deceleration applies to all molecules regardless of their arrival time, and thus is effective for continuous (or long-pulsed) molecular beams.

Our apparatus has two deceleration stages with $\Bmax \approx 2.8$~T, leading to $\DEstage \approx 3.8$~K. The total energy removal for two stages ($\sim$7.6~K) is therefore well matched to CBGBs, which can have typical kinetic energies $E_\text{kin} \lesssim 8$~K~\cite{hutzler2012buffer,patterson2007bright}. Because a fixed \textit{energy} is removed in each stage, the decelerator will slow to rest \red{$1\muB$} molecules of any mass produced at or below the same threshold temperature~\cite{Note1}. \red{The energy removal capability scales approximately linearly with ground-state magnetic moment.}  Despite the overhead associated with superconducting coils (principally the use of a cryocooler), this setup leads to several technical advantages over permanent magnets, including larger $\Delta E_\text{stage}$, increased transverse acceptance, and excellent vacuum due to cryopumping. Furthermore, it is straightforward to introduce optical access transverse to the molecular beam so that laser-molecule interaction occurs only at the desired points.

To begin an experimental cycle, CaOH molecules are produced by laser ablation of Ca(OH)$_2$ inside a buffer-gas cell held at 2~K. [See Fig.~\ref{fig:Setup}(b) for a schematic overview of the experimental setup.] The CaOH molecules thermalize with $^3$He buffer-gas that fills the cryogenic cell at densities around $10^{15}$~cm$^{-3}$. Molecules are extracted through a 7~mm aperture and then enter a second cell, 40~mm long and separated from the first cell by a 2.5~mm gap, that reduces the forward-velocity boosting associated with molecular beam extraction~\cite{hutzler2012buffer, Barry2011}. The second cell is thermally isolated from the first cell and is cooled to 0.7~K by a pumped $^3$He refrigerator. Molecular beams of CaOH are produced with forward velocities in the range $v_f \sim 45-70$~m/s ($E_\text{kin} \sim 7-16$~K), tunable by a combination of ablation energy and buffer-gas density.

The molecular beam propagates and then enters a separate cryogenic region [``magnet region" in Fig.~\ref{fig:Setup}(b)] containing two identical superconducting solenoids. (See \cite{Note1} for details about the magnet design.) Slots at the center of each solenoid provide optical access transverse to the molecular beam in two high-field pumping regions (D1 and D3). A gap between the two solenoids provides similar access in the low-field region (D2). Two lasers are used for the deceleration steps, both addressing molecules in the $N''=1$ rotational state on the $\APiHalf(000) \leftarrow\XSigma(000)$ $P$-branch transition~\cite{Stuhl2008}. A single laser beam detuned $\mysim38$~GHz below the zero-field resonance is split to provide up to 250~mW to each of D1 and D3. \red{These laser beams are cylindrically expanded to cover the transverse extent of the molecular beam and fill the gap at the center of each solenoid (Gaussian diameter $\approx 25 \times 12$~mm). Razor blade collimators prevent light from scattering on the coil forms.} A separate laser, detuned $\mysim1.4$~GHz above the zero-field resonance, provides 200~mW to D2 \red{(spherically expanded to a Gaussian diameter $\approx 25$~mm).} \red{All laser beams are polarized perpendicular to the magnetic field direction.}

\begin{figure}[tb]
    \centering 
    \includegraphics[width=0.95\columnwidth]{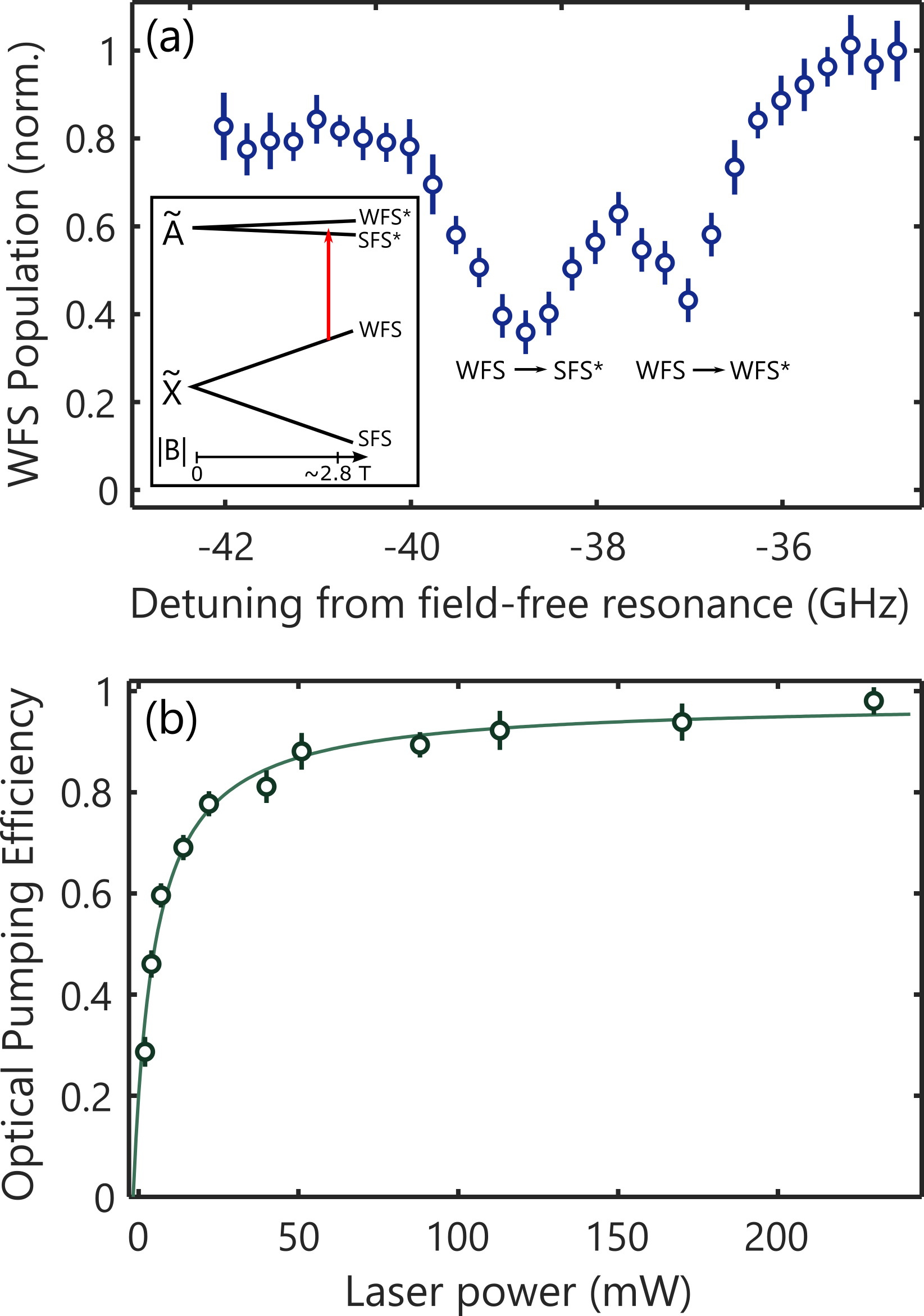}
    \caption{Optical pumping measurements in the first high-field deceleration pumping region, D1. (a) WFS state population as a function of laser frequency applied in D1 with $\Bmax=2.8$~T. The doublet arises due to Zeeman splitting in the $\APiHalf(000, J^\prime=1/2)$ level. The offset at lower frequencies is due to asymmetrical Zeeman broadening present in the pumping region. Inset: Level structure giving rise to the observed transitions. (b) Efficiency of pumping from WFS to SFS when both features of the observed doublet are driven simultaneously. \red{The solid line is a fit to a saturation model indicating asymptotic pumping efficiency of 97$^{+3}_{-2}\%$ and saturation power of $8(3)$~mW.}}
    \label{fig:OpticalPumping}
\end{figure}

A key feature of Z-S deceleration is the possibility of high-efficiency optical pumping using only a few photon cycles and relatively low laser power; this can lead to very efficient slowing. We studied the optical pumping performance by monitoring the population in WFS states downstream while applying various combinations of optical pumping light in D1, D2, and D3. Figure~\ref{fig:OpticalPumping}(a) shows the optical pumping efficiency from WFS to SFS states when $\myB \approx2.8$~T and 150~mW of optical pumping light is applied in D1. Two optical pumping features are observed due to excited-state Zeeman splitting~\cite{Note1}. Each ground-state Zeeman sublevel couples to a single excited-state sublevel; \red{to address all the WFS sublevels, optical excitation through both excited state features is required due to the details of angular momentum selection rules in the fully decoupled regime~\cite{Fitch2016}.} To do so, the optical pumping light is passed through a resonant electro-optical modulator tuned to $f_\text{EOM} \approx 1.7$~GHz with \red{$\beta \approx 1.7$}. In Fig.~\ref{fig:OpticalPumping}(b), we show the optical pumping efficiency when both Zeeman features are addressed in D1. At laser powers above about 200~mW, we observe $\mygtr97\%$ optical pumping efficiency. Repeating these measurements in D3 yields identical results. In D2, the excited-state Zeeman shifts are unresolved, so full optical pumping can be achieved with a single frequency. We find optical pumping efficiencies $\mygtr95\%$ in D2 with $\mysim150$~mW of laser power.

Following the solenoids, the molecules propagate for a distance of 35~cm through an octupole magnetic guide, entering a room-temperature vacuum chamber. A state-preparation region with 150~mW of laser power, S2 in Fig.~\ref{fig:Setup}(b), is used to return population to the WFS manifold just before the magnetic guide. In the detection region, the molecules are detected via laser-induced fluorescence by driving the $\APiHalf(000, J'=1/2^+) \leftarrow \XSigma(000,N''=1^-)$ (626~nm) and $\BSigma(000,N'=0^+) \leftarrow \XSigma(100,N''=1^-)$ (574~nm) transitions in a Doppler-sensitive configuration. The resulting $\BSigma(000) \rightarrow \XSigma(000)$ fluorescence at 555~nm allows nearly background-free detection. A velocity resolution of $\mysim4$~m/s is typical. Based on independent calibration of the fluorescence detection system, we estimate that, to within a factor of two, the full molecular beam contains $1.5\times10^5$ molecules per pulse in the detection region.

\begin{figure}[tb]
    \centering 
    \includegraphics[width=1\columnwidth]{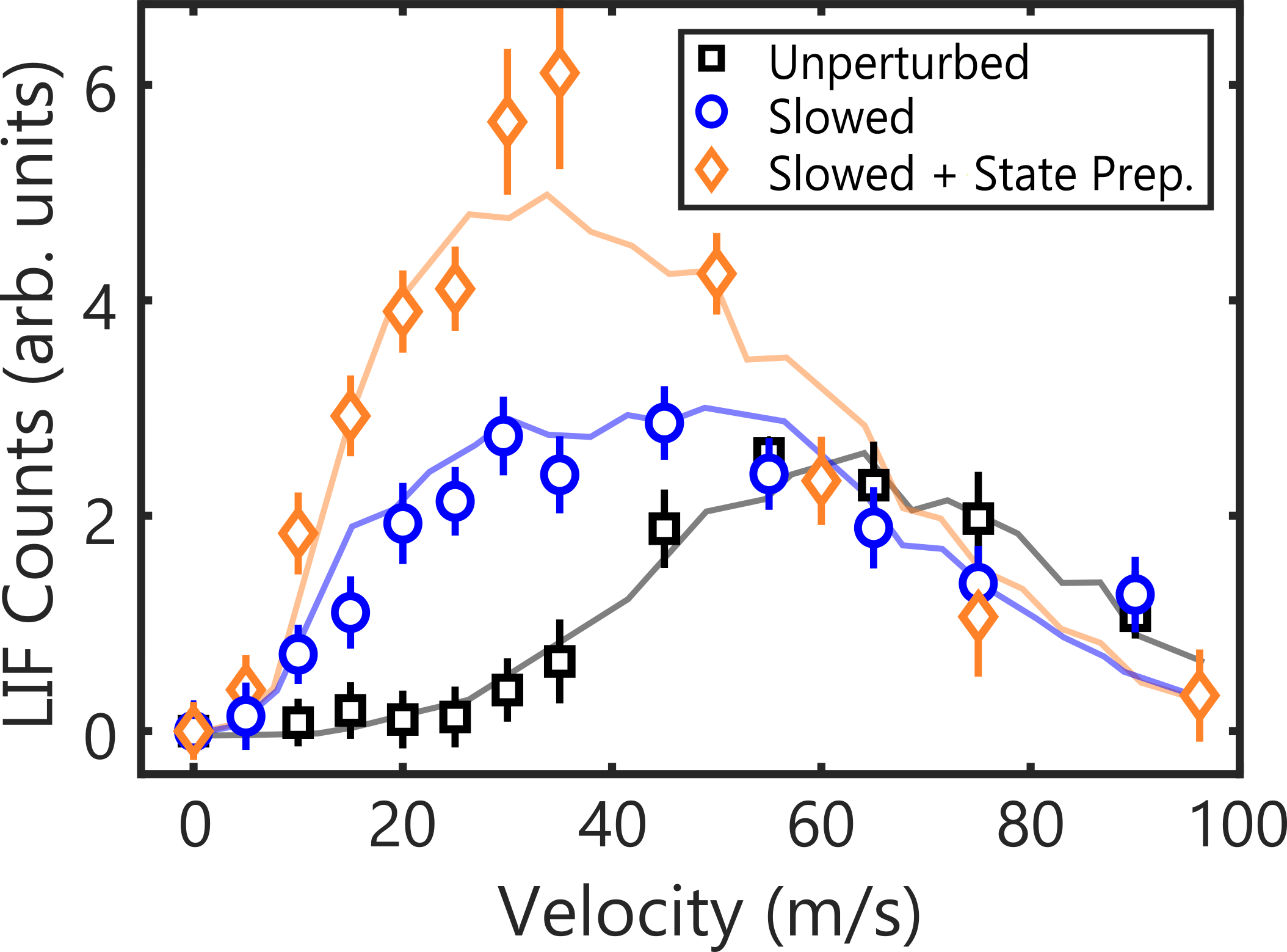}
    \caption{CaOH velocity distributions in the detection region with (blue circles) and without (black squares) Zeeman-Sisyphus deceleration applied. Also shown (orange diamonds) is the velocity distribution when all molecules are prepared in weak-field-seeking states prior to slowing (using pumping region S1). Solid lines are the results of Monte Carlo trajectory simulations.}
    \label{fig:VelocityDistribution}
\end{figure}

The performance of the decelerator is shown in Fig.~\ref{fig:VelocityDistribution}, which presents the integrated laser-induced fluorescence as a function of velocity. We also show in solid lines the results of Monte Carlo trajectory simulations. In the unperturbed molecular beam, the population with velocities below 20~m/s is consistent with zero; we estimate an upper bound of $0.2\%$ of the population in this range. Fewer than $0.02\%$ of the molecules have velocities below 10~m/s. When the decelerator is turned on, we observe a shift in the peak velocity and a dramatic enhancement of the low-velocity tail. Due to order-of-unity transverse focusing effects within the magnets, more molecules are detected downstream in the slowing configuration as compared to the unperturbed beam. Such focusing is reflected accurately in our simulations. 

For the slowed beam, the fractions of molecules (relative to the unperturbed population) below 20~m/s and 10~m/s increase to $9(2)\%$ and $1.5(3)\%$, respectively. We can further increase the number of slow molecules by preparing all molecules in the WFS manifold before entering the Z-S magnets [region S1 in Fig.~\ref{fig:Setup}(b)]. When this is done, the fraction of slow molecules rises to $24(3)\%$ below 20~m/s and $3.5(5)\%$ below 10~m/s. The fraction of slow molecules is therefore enhanced by at least two orders of magnitude following deceleration. Figure~\ref{fig:VelocityDistribution} also shows the results of Monte Carlo simulations that take as input experimentally measured laser parameters and accurate, three-dimensional magnetic field profiles for both the superconducting coils and the magnetic guide. We find excellent agreement between the simulations and experimental results, indicating that the details of the slowing process are modeled accurately.

\begin{figure}[tb]
    \centering 
    \includegraphics[width=0.98\columnwidth]{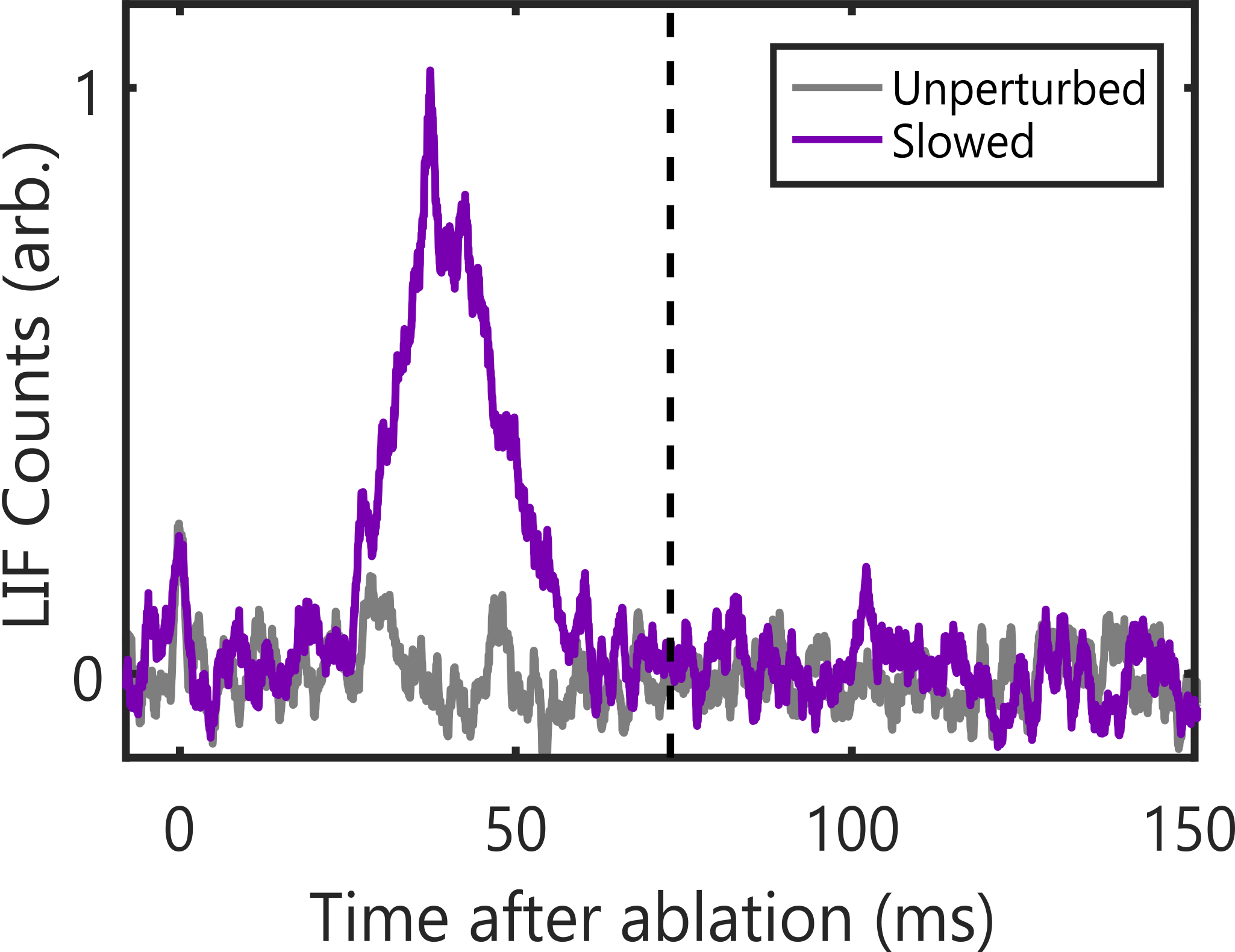}
    \caption{Laser-induced fluorescence due to molecules in the $15$~m/s velocity class without (gray) and with (purple) the optical pumping lasers applied. The dashed vertical line indicates the earliest possible time-of-arrival for molecules initially produced at 15~m/s.}
    \label{fig:LowVelocityClass}
\end{figure}

Many molecular traps (e.g., electrostatic~\cite{prehn2016optoelectrical}, magnetic~\cite{Sawyer2008Molecular}, magneto-optical~\cite{Tarbutt2015b,Williams2017}, and microwave~\cite{DeMille2004Microwave} traps) can capture molecules with kinetic energy below about 1~K, which for CaOH is equivalent to $v_\text{trap} \sim 15$~m/s. Figure~\ref{fig:LowVelocityClass} shows the time-dependent laser-induced fluorescence resulting from molecules with forward velocity in the range $13-17$~m/s. With the Z-S slowing applied, clear accumulation is observed in this velocity class. The peak laser-induced fluorescence occurs at $\Delta t_\text{max} \approx42$~ms after ablation, well before the earliest possible arrival time for molecules produced with such velocities ($\Delta t_\text{natural}\approx 75$~ms). This provides additional confirmation that these molecules were decelerated from much higher velocities. The observations also agree well with the results of our numerical simulations, which predict that molecules produced near 50~m/s are slowed to around 15~m/s and arrive in the detection region 40~ms after the ablation pulse.

To determine the number of photons scattered by the slowed molecules, we measure the population pumped into the \XSigma$(100)$ level. Based on the known vibrational branching ratios~\cite{Zhang2021Accurate}, these measurements are consistent with an average of 1.3(2) photons scattered in each optical pumping step~\footnote{The number is greater than 1 because each photon scatter flips the electron spin with less than unit probability.}. The relative rotational branching ratios to ground-state SFS and WFS manifolds predict that an average of 1.5~photon scatters are required to achieve state transfer, in good agreement with the measured value. The slowing process therefore requires approximately 4.5 photon scatters per molecule, while the two state-preparation steps use an additional 3 photons per molecule. That is, $\mysim7$ photon scatters are sufficient to decelerate a CaOH molecule near the peak of the distribution by $\Delta v_f \approx 35$~m/s. By contrast, the radiative force due to 7 scattered photons would slow a CaOH molecule by $\Delta v_f \approx 0.1$~m/s.

In summary, we have demonstrated a general method to decelerate paramagnetic molecules to trappable velocities, requiring just a few optical pumping steps. Using CaOH as a test species, we observed the production of molecules with velocities below 15~m/s, enhancing this slow population by at least two orders of magnitude. The measured velocity and time-of-arrival distributions are in excellent agreement with numerical simulations. After slowing, approximately $3\times10^4$ molecules per pulse are found in velocity classes capturable by long-lived traps (e.g., MOT or magnetic). The decelerated molecules are detected in a room-temperature, ultra-high vacuum environment that is ideally suited for subsequent trapping and deep laser cooling. The flux of slow molecules can be increased in several ways, including laser-enhanced molecule production~\cite{Jadbabaie2020}, shortening the beamline, and adding a magnetic lens after the CBGB cell. Using previously demonstrated molecular fluxes~\cite{Jadbabaie2020, Zhang2021Accurate}, these changes could increase the number of slow CaOH molecules by $\mysim 100\times$. 

While we have focused in this work on the linear molecule CaOH, the energy removal demonstrated here depends only on ground-state Zeeman shifts and can be applied generically to other paramagnetic species capable of optically cycling at least a small number of photons. As we have shown, the required laser technology is straightforward and requires around three lasers producing $\lesssim$1~W each. Our results therefore open up a wide range of future directions that rely on slow fluxes of polyatomic molecules. For small molecules, including CaOH, SrOH, and YbOH, this work establishes a path to load polyatomic molecular magneto-optical traps with numbers comparable to those demonstrated previously for some diatomic molecules~\cite{norrgard2015sub, Collopy2018}. \red{For molecules without convenient optical cycling transitions, it is possible that microwave transitions within the ground state could be used to realize the spin flip required for deceleration, although the state transfer efficiency may be lower in such a scheme.} The low output velocities of the decelerator are well matched to the capture velocities of certain conservative traps, potentially enabling direct trap loading~\cite{lu2014magnetic} of complex polyatomic molecules.

\begin{acknowledgments}
We gratefully acknowledge helpful discussions with the PolyEDM collaboration, including the Hutzler group (Caltech), Timothy C. Steimle, and Amar Vutha. This work was supported by the Heising-Simons Foundation, the Gordon and Betty Moore Foundation, the Alfred P. Sloan Foundation, and the AFOSR. Z.D.L. was supported by the Center for Fundamental Physics (Fundamental Physics Grant) and the Templeton Foundation. H.S. acknowledges financial support from the Ezoe Memorial Recruit Foundation, and N.B.V. acknowledges financial support from the NDSEG fellowship.
\end{acknowledgments}

\typeout{}
\bibliography{ZS_library}

\end{document}